\documentclass[aps,prl,twocolumn,superscriptaddress,nofootinbib,longbibliography]{revtex4-2}

\usepackage{amsmath,amssymb,amsfonts,mathtools,bm}
\usepackage{graphicx}
\usepackage{hyperref}
\usepackage{xcolor}
\usepackage{physics}
\usepackage{bbm}

\begin{document}

%\title{Exact Bulk-Boundary Pairs Beyond the Asymptotic Dictionary in AdS/CFT}
\title{Exact Bulk-Boundary Pairs in AdS/CFT}

\author{Xin Jiang}
\email{domoki@stu.scu.edu.cn}
\affiliation{College of Physics, Sichuan University, Chengdu, 610065, China}

\author{Peng Wang}
\email{pengw@scu.edu.cn}
\affiliation{College of Physics, Sichuan University, Chengdu, 610065, China}

\author{Haitang Yang}
\email{hyanga@scu.edu.cn}
\affiliation{College of Physics, Sichuan University, Chengdu, 610065, China}

\begin{abstract}
We show that for a CFT$_D$ on a flat open solid torus, the two point function in the Weyl
frame is exactly paired with a finite geodesic lying entirely in the AdS$_{D+1}$ bulk interior. This relation
is exact and  requires neither large $N$, strong coupling, nor heavy operators. 
The exactness is that of conformal kinematics; no semiclassical bulk dynamics is assumed.
The standard boundary-anchored relation is a singular limit of the exact pair.
For the free scalar, a mode expansion along $S^1$ generates an infinite tower of effective masses 
on $H_{D-1}$, whose intricate propagators resum exactly to the same simple higher-dimensional geodesic expression. 
Together with another exact pair
between disjoint entanglement entropy and entanglement wedge cross-section found on the same open
solid torus, this result points toward a broader exact-pair program in AdS/CFT.
\end{abstract}

%With another exact pair
%between disjoint entanglement entropy and entanglement wedge cross-section found on the same open
%solid torus, it thus suggests an exact-pair program in AdS/CFT.

%Since the scalar bulk-to-bulk propagator on AdS$_{D+1}$ is fixed by the geodesic invariant and the conformal dimension, 
%the construction also yields an exact reconstruction of the corresponding bulk propagator. 

%We present an exact bulk-boundary pair in AdS/CFT and propose a broader exact-pair program. 
%For a CFT$_D$ on a flat open solid torus, the two point function in the Weyl frame is exactly
%paired with a finite geodesic lying deep in the bulk interior. 
%Unlike the usual geodesic picture, this relation is exact and requires neither a large-$N$ limit nor heavy operators. 
%It thus provides a second example of an exact pair, in addition to the previously identified relation between disjoint entanglement entropy and entanglement wedge cross section. 
%For the free scalar, a mode expansion further shows how the distinguished $S^1$ generates effective masses in the remaining dimensions, whose intricate propagators resum to the same simple geodesic expression.

\maketitle

\paragraph*{Introduction.}
A central challenge in holography is to identify boundary observables that encode bulk interior geometry {\it exactly}. 
Most familiar entries in the AdS/CFT dictionary instead relate CFT observables to bulk quantities anchored at the asymptotic boundary \cite{Maldacena:1997re,Gubser:1998bc,Witten:1998qj,Ryu:2006bv,Hubeny:2007xt}. Their geometric partners are divergent 
and require cutoff-dependent subtraction. 
This asymptotic pairing obscures a sharper possibility: 
finite geometric data deep in the bulk may admit direct exact encoding in properly organized CFT observables.
In other words, are there exact bulk-boundary pairs in AdS/CFT, valid beyond the usual semiclassical regime?

%One of the central lessons of holography is that geometric quantities in asymptotically anti-de Sitter space encode field-theoretic observables.
%Most familiar examples, however, are either semiclassical, large-$N$, or restricted to special operator sectors.
%This raises a natural question: are there exact bulk-boundary pairs in AdS/CFT, valid beyond the usual semiclassical regime?

This issue is especially important for bulk reconstruction. 
If the interior is accessed only through boundary-attached, regularized quantities, 
reconstruction is necessarily indirect and difficult. 
A qualitatively different route would be to identify exact boundary observables 
dual to finite bulk quantities already living inside bulk. 

\begin{figure}[h]
\centering
\includegraphics[scale=0.35]{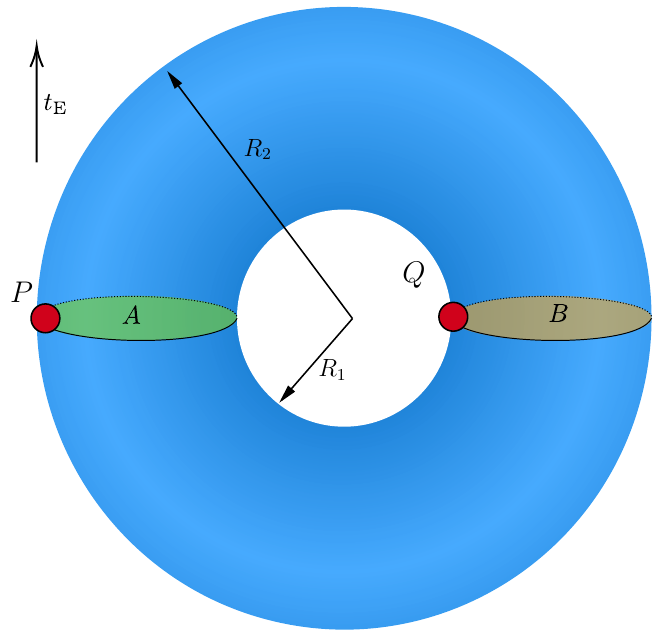}
\caption{The CFT$_D$ lives on the flat open solid torus $S^1\times \mathbb{B}^{D-1}$ with  Euclidean flat metric.
The Euclidean time direction is upward. This is a pure state with disjoint subregions $A$ and $B$. 
The disjoint entanglement entropy $S_{\rm disj}(A:B)$ is finite and exactly matches the bulk EWCS $E_W$ of $A$ and $B$.
\label{fig:solid-torus}}
\end{figure}

\begin{figure}[h]
\centering
\includegraphics[scale=0.33]{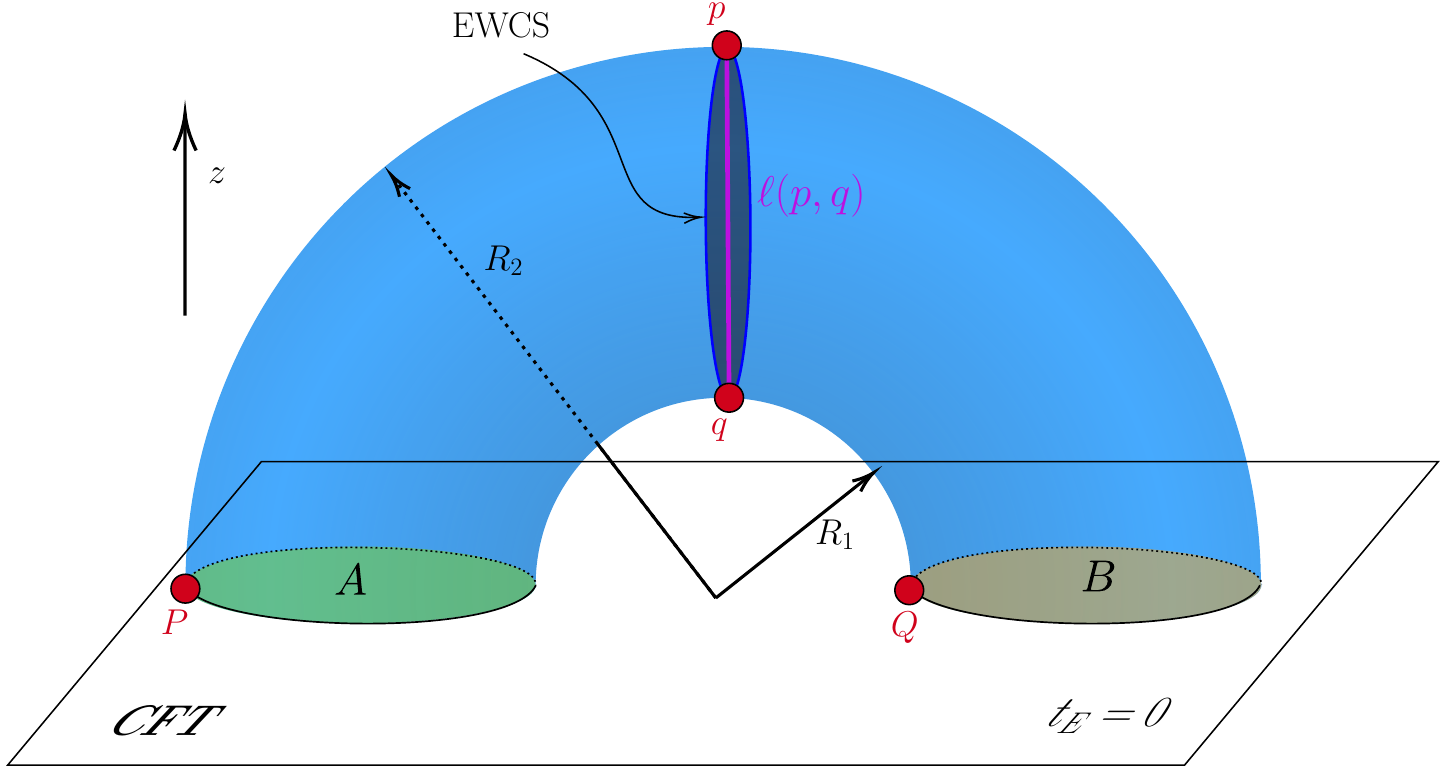}
\caption{AdS$_{D+1}$ geometry $ds_{EAdS}^2 = \frac{dt_E^2 + dz^2 +\Sigma dx_I^2}{z^2}$ with the bulk direction $z$ upward. 
%$A$ and $B$ are cross sections of the boundary $D-$dim open solid torus. 
$P$ and $Q$ are the correlator points in CFT$_D$.
The shaded sphere denotes the EWCS of $A$ and $B$. 
$p$ and $q$ are antipodal points of the EWCS.
The geodesic length between $p$ and $q$ is $\ell(p,q) = \log\frac{R_2}{R_1}$.} 
\label{fig:RT}
\end{figure}

A first affirmative example was identified in our previous work \cite{Jiang:2025jnk}, 
\[
S_{\rm disj}(A:B) \equiv E_W,
\]
where $S_{\rm disj}(A:B)$ is the von Neumann entropy between {\it disjoint complementary}
spacelike regions $A$ and $B$ in CFT$_D$ living on an open solide torus, 
as illustrated in Fig. \ref{fig:solid-torus}.
$E_W$ is the entanglement wedge cross section (EWCS) of $A$ and $B$, 
which is entirely in the bulk interior, as shown by the shaded sphere in Fig. \ref{fig:RT}. 

In this work we provide a 
second example and use it to suggest a broader exact-pair program in AdS/CFT.
We consider a CFT$_D$ placed on an open solid torus and study its two point function.
In the Weyl frame, we find that the correlator is exactly paired with 
a finite geodesic segment lying entirely in the bulk interior, 
identified as the antipodal geodesic on the EWCS.
This relation is exact, finite, and does not rely on large $N$, strong coupling, or heavy operators.
The standard boundary-anchored relation is a singular limit of this exact pair.

In addition to its conceptual significance, 
the construction has an illuminating microscopic explanation in the free scalar CFT.
There, a mode expansion along the distinguished $S^1$ produces an infinite 
tower of effective masses on the remaining $(D-1)$ dimensions.
The resulting $(D-1)$-dim  massive propagators are individually very intricate, 
but their sum collapses to the remarkably simple $(D+1)$-dim finite geodesic expression.
%Weyl frame correlator
%of the open solid torus. 
%This provides an independent and nontrivial check of the exact pairing.

Our result should be viewed not merely as an isolated identity, 
but as evidence for a wider organizing principle:
certain CFT observables may admit exact geometric partners in 
the bulk even outside the usual large-$N$ or saddle-point framework.
The relevant dual objects are not the bare planar quantities alone, but the triplet
({\it CFT, state, boundary geometry}).

Since the scalar bulk-bulk propagator on AdS$_{D+1}$ is fixed by the geodesic invariant and the conformal dimension, 
the construction also yields an exact reconstruction of the corresponding bulk propagator.

\paragraph*{Setup: CFT$_D$ on an open solid torus.}
Our setting is Euclidean throughout.
For a CFT in $D$-dim  spacetime with flat metric
$\mathrm{d}s_E^{2}=\mathrm{d}t_{\text{E}}^{2}+\mathrm{d}y^{2}+  \underset{I=1}{\overset{D-2}{\sum}}
\mathrm{d}x_{I}^{2}$,
we consider quantum fields living in  an open solid torus $S^1\times \mathbb{B}^{D-1}$:
%\begin{widetext}
%\begin{equation}
\begin{equation}
\mathcal{B}_D := \left\{ \left(\sqrt{t_{\text{E}}^{2} + y^{2}} - \frac{R_{2} + R_{1}}{2}\right)^{2} + \sum_{I=1}^{D-2} x_{I}^{2} < \left(\frac{R_{2} - R_{1}}{2}\right)^{2} \right\},
\end{equation}
%\end{equation}
%\end{widetext}
with $R_{2}>R_{1}>0$   as shown in Fig. \ref{fig:solid-torus}. 
$\mathbb{B}^{D-1}$ denotes a $(D-1)$-dim open ball.
Note since the solid torus is open, it is flat. 

Working on this open solid torus, in \cite{Jiang:2025jnk}, we calculated the von Neumann 
entropy $S_{\mathrm{disj}}(A:B)$ between  $A$ and $B$ on $t_E=0$ time slice for all dimensions. 
$S_{\mathrm{disj}}(A:B)$ is indeed finite. 
We showed the exact pair $S_{\mathrm{disj}}(A:B) \equiv E_W$. 
Under the adjacent limit, this finite disjoint entanglement entropy 
reproduces the usual divergent entanglement entropies.
We now demonstrate that this open solid torus also leads to 
another exact pair: CFT correlator/Bulk geodesics.

Making the coordinate transformations, $t_{\text{E}}=r\sin\theta$,
$y=r\cos\theta$, after a
Weyl rescaling $\mathrm{d}s_{E}^{2}\rightarrow\mathrm{d}s_{E}^{2}/r^{2}$,
the spacetime metric in the Weyl frame is
$\mathrm{d}s_{\mathcal{W}}^{2}=\mathrm{d}\theta^{2}+\left(\mathrm{d}r^{2}+\sum_{I=1}^{D-2} 
\mathrm{d}x_{I}^{2}\right)/r^{2}$.
Select two points $P(r=R_2, \theta_P=\pi, x_I=0)$ and 
$Q(r=R_1, \theta_Q=0, x_I=0)$ 
as shown in Fig. \ref{fig:solid-torus}.
For a primary scalar operator $\mathcal{O}$ of dimension $\Delta$,
it is straightforward to calculate the two-point function in the Weyl frame,

\begin{eqnarray}
G_{\mathcal{W}}(P,Q) &:=& 
\langle \mathcal{O}(P)\mathcal{O}(Q)\rangle_{\mathcal{W}} \nonumber\\
&=& R_2^\Delta R_1^\Delta 
\frac{C_\Delta}{|R_1^2+R_2^2 - 2R_1 R_2 \cos\delta\theta|^\Delta}\nonumber\\
&=& \frac{C_\Delta}{2^\Delta\, |\cosh L(P,Q) -  \cos\delta\theta|^\Delta},
\label{eq:WithTheta}
\end{eqnarray}
where $C_\Delta$ is a normalization constant and the distance is
\begin{equation}
 L(P,Q)\equiv \log\frac{R_2}{R_1}.
 \label{eq:vGL}
\end{equation} 
For $P$, $Q$ selected as in Fig. \ref{fig:solid-torus}, $\delta\theta=\pi$,
\begin{equation}
G_{\mathcal{W}}(P,Q) = \frac{C_\Delta}{\Big[2\cosh\frac{L(P,Q)}{2}\Big]^{2\Delta}}.
\end{equation}

%We work in Euclidean signature.
%Let the CFT live on a flat solid-torus geometry
%\begin{equation}
%\mathcal{M}_d \simeq S^1 \times B^{d-1},
%\end{equation}
%with metric
%\begin{equation}
%ds^2_{\mathcal{M}} = [\,\text{insert explicit flat solid-torus metric}\,].
%\end{equation}
%Here the $S^1$ factor is distinguished, while the remaining directions form a $(d-1)$-dimensional ball with appropriate identifications so that the full geometry is smooth and has no physical boundary in the sense relevant for the CFT construction.
%Equivalently, if your preferred notation is different, replace the above by the explicit coordinate system used in your derivation.

%A key role is played by a Weyl transformation
%\begin{equation}
%ds^2_{\mathcal{M}} = \Omega^2(x)\, ds^2_{\rm W},
%\end{equation}
%to a frame in which the two-point function takes a particularly transparent form.
%We refer to $ds^2_{\rm W}$ as the natural Weyl frame of the problem.
%For a primary scalar operator $\mathcal{O}$ of dimension $\Delta$, the correlator transforms as
%\begin{equation}
%\langle \mathcal{O}(x_1)\mathcal{O}(x_2)\rangle_{\mathcal{M}}
%=
%\Omega(x_1)^{-\Delta}\Omega(x_2)^{-\Delta}
%\langle \mathcal{O}(x_1)\mathcal{O}(x_2)\rangle_{\rm W}.
%\label{eq:Weyl-transform}
%\end{equation}
%The central statement of this paper concerns the exact form of the correlator in this Weyl frame.

\paragraph*{Exact pairing: finite geodesic and correlator.}

It is easy to recognize that $L(P,Q)$ in eq. (\ref{eq:vGL}) is nothing but a
vertical geodesic length in AdS geometry. To see this, 
referring to Fig. \ref{fig:RT}, in the EAdS$_{D+1}$,
setting the radius to be the unity, 
$ds_{EAdS}^2 = \frac{dt_E^2 + dz^2 +\Sigma dx_I^2}{z^2}$,
the $D-$dim open solid torus 
locates in the $R^D$ which is the boundary of EAdS$_{D+1}$.  
%Spheres $A$ and $B$ are the cross sections of the  solid torus.
The shaded sphere is the EWCS of $A$ and $B$. 
The geodesic length between the antipodal point $p$ and $q$ is
\begin{equation}
 \ell(p,q)= \int^{R_2}_{R_1} \frac{dz}{z}= \log\frac{R_2}{R_1},
 \label{eq:EWCS}
\end{equation}
which is exactly equal to the distance $L(P,Q)$ in  eq. (\ref{eq:vGL}).

\begin{figure}[h]
\centering
\includegraphics[scale=0.5]{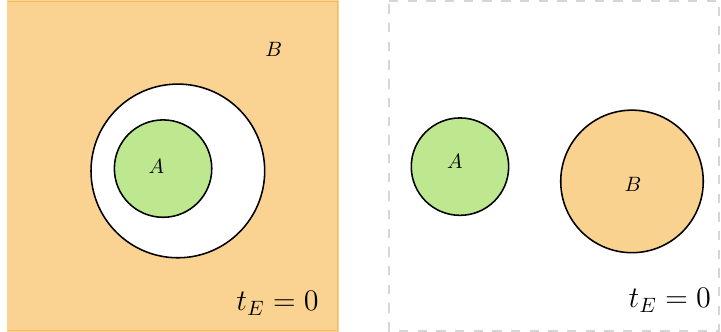}
\caption{$t_E =0$ slice (colored regions)  of the solid torus.
Left panel:the cavity configuration.
Right panel: the juxtaposed configuration.}
\label{fig:two-config}
\end{figure}

This perfect agreement certainly also applies to arbitrary spacelike geodesics 
and correlators of asymmetric configurations.
Generally, there are two different  configurations of solid torus,
namely, the cavity configuration and the juxtaposed configuration as illustrated in Fig. \ref{fig:two-config}.
For two spheres with centers $\vec x$ and $\vec x'$, radii $r$ and $r'$, there is a conformal invariant quantity, 
the inversive product  \citep{beardon2012geometry}
\begin{equation}
\varrho=\left|\frac{r^{2}+r^{\prime2}-\vert\vec{x}-\vec{x}^{\prime}\vert^{2}}{2rr^{\prime}}\right|.
\label{eq:inversive product}
\end{equation}
The inversive product $\varrho$ 
remains invariant under global conformal transformations. 
Since both the Weyl frame correlator and the corresponding bulk geodesic length depend only on $\varrho$, it is sufficient to establish the relation in a convenient symmetric representative of each conformal class. The exact pair therefore extends from the symmetric solid torus to arbitrary spacelike configurations, including both the cavity and juxtaposed cases.
So, it is more universal to write the distance (\ref{eq:vGL}) and (\ref{eq:EWCS})
in terms of $\varrho$.
Since for the solid torus in Fig. \ref{fig:solid-torus},
$\frac{R_2}{R_1}=\frac{\sqrt{\varrho+1}+\sqrt 2}{\sqrt{\varrho+1}-\sqrt 2}$, we have
\begin{equation}
\boxed{L(P,Q)=\ell(p,q)=\log \frac{\sqrt{\varrho+1}+\sqrt 2}{\sqrt{\varrho+1}-\sqrt 2}.}
\label{eq:GL-rho}
\end{equation}
Note for simplicity, in the above derivations, we set $\delta x_I = x_I(P) -x_I(Q) =0$. 
It is not hard to see that this exact pair also holds for $\delta x_I \not=0$. 
In this case,
the bulk geodesic is not the vertical antipodal one anymore, but shifts accordingly with $\delta  x_I^2$.

%So, for an arbitrary  spacelike geodesic $\gamma$ which belongs to the conformal class $C_\gamma$,
%we  conformally transform it to the vertical geodesic $\ell_\gamma$ in $C_\gamma$,
%and then apply the inverse map to the symmetric solid torus in $C_\gamma$ to obtain the 
%dual CFT configuration of $\gamma$. 

\begin{figure}[h]
\centering
\includegraphics[scale=0.38]{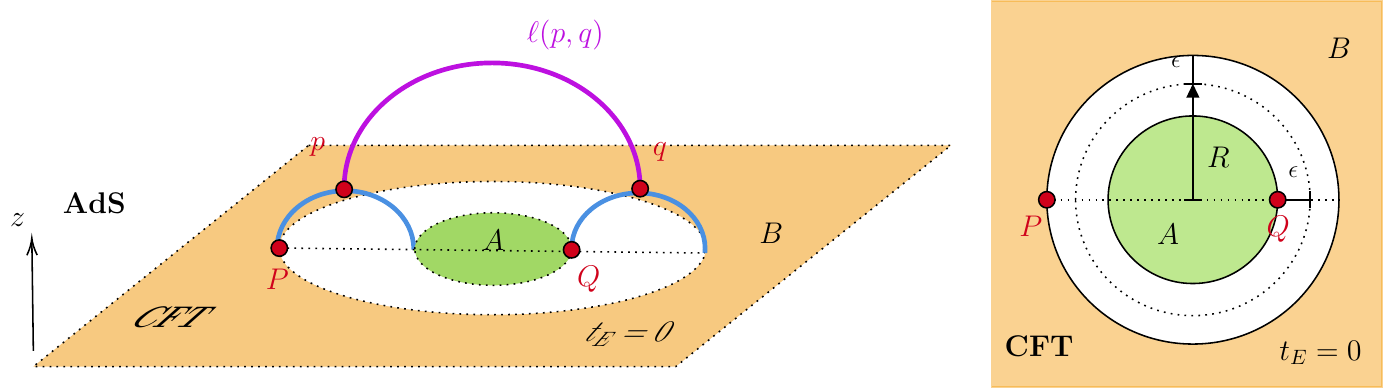}
\caption{The cavity configuration, the dual bulk geodesic is the purple curve $\ell(p,q)$.
We should use the conformal invariant expression (\ref{eq:GL-rho}) to compute $\ell(p,q)$
since eq. (\ref{eq:vGL}) only applies
to the symmetric solid torus.}
\label{fig:limit}
\end{figure}

We emphasize that the dual bulk geodesic is determined by the correlator in the Weyl frame, not by the planar correlator alone.
The Weyl frame correlator is determined by both the planar correlator and 
the configuration of solid torus.
So, for different solid tori, the same planar correlators lead to different 
Weyl frame correlators and consequently different dual bulk geodesics.
The relevant datum is therefore the triplet: ({\it CFT, state, boundary geometry}).
This is mostly
manifested in the cavity configuration in Fig.
(\ref{fig:limit}). There, although the locations of $P$ and $Q$ are the same 
as those in the symmetric solid torus in Fig. \ref{fig:RT}, 
the dual bulk geodesics are totally different from the vertical one in Fig. \ref{fig:RT}.

The identification of the dual geodesic is simplest when 
one starts from a given solid torus configuration, 
either the cavity configuration or the juxtaposed configuration. 
The dual bulk geodesic is simply
the antipodal geodesic in the EWCS.

We therefore obtain the exact bulk-boundary pair:
\begin{equation}
\boxed{G_{\mathcal{W}}(P,Q) = \frac{C_\Delta}{\Big[2\cosh\frac{\ell(p,q)}{2}\Big]^{2\Delta}},}
\end{equation}
where $P$,$Q$ are located on the inner or outer radius of the solid torus with an angle difference $\pi$.
$\ell(p,q)$ taking the form of eq. (\ref{eq:GL-rho}), is the length of the 
antipodal geodesic in the EWCS of the (a)symmetric solid torus.

%
%
%Our main result is that the Weyl-frame two-point function is exactly paired with a finite geodesic quantity in the bulk.
%More precisely,
%\begin{equation}
%\langle \mathcal{O}(x_1)\mathcal{O}(x_2)\rangle_{\rm W}
%=
%\mathcal{N}_\Delta \,
%\mathcal{F}\!\left(\ell_{\rm bulk}(x_1,x_2)\right),
%\label{eq:main-pairing-general}
%\end{equation}
%where $\ell_{\rm bulk}(x_1,x_2)$ denotes the length of a finite geodesic lying deep in the Euclidean bulk interior, and $\mathcal{F}$ is the explicit function determined in our construction.
%In the simplest form encountered in our examples, Eq.~\eqref{eq:main-pairing-general} reduces to
%\begin{equation}
%\langle \mathcal{O}(x_1)\mathcal{O}(x_2)\rangle_{\rm W}
%=
%\mathcal{N}_\Delta \,
%\e^{-\Delta \,\ell_{\rm bulk}(x_1,x_2)},
%\label{eq:main-pairing-exp}
%\end{equation}
%or equivalently
%\begin{equation}
%-\frac{1}{\Delta}\log
%\frac{\langle \mathcal{O}(x_1)\mathcal{O}(x_2)\rangle_{\rm W}}{\mathcal{N}_\Delta}
%=
%\ell_{\rm bulk}(x_1,x_2),
%\label{eq:main-pairing-log}
%\end{equation}
%depending on normalization conventions.
%If your exact formula differs by additive constants or by a known kinematic prefactor, replace Eqs.~\eqref{eq:main-pairing-exp}--\eqref{eq:main-pairing-log} with the precise statement.

Several features deserve emphasis.
First, $\ell(p,q)$ is a finite  geodesic lying entirely in the bulk interior; 
it is not a boundary-anchored geodesic and requires no renormalization.
Second, the relation is exact rather than semiclassical; 
neither large $N$ nor heavy operators (large $\Delta$) is needed.
It is therefore qualitatively different from the standard geodesic 
approximation for heavy operators in holographic large-$N$ CFTs.

%Since the scalar bulk-bulk propagator on AdS$_{D+1}$ is fixed by the geodesic  and the conformal dimension, 
%the construction also yields an exact reconstruction of the  bulk propagator. 

\paragraph{Kinematic nature of the exactness.}
Let us stress what is, and what is not, meant by the exactness of the pair.
The correlator of a scalar primary is fixed by conformal covariance. After the
solid torus Weyl transformation, the ordinary flat space conformal invariant is rewritten as
a finite hyperbolic invariant, which is then represented by the length of a geodesic in an
associated AdS geometry. Thus the exactness established here is the exactness of conformal
kinematics. It does not rely on, nor does it claim by itself, the existence of a semiclassical
Einstein bulk. For a generic CFT, the AdS geometry appearing in this construction should be
viewed as a kinematic or auxiliary AdS geometry associated with the solid torus Weyl frame.
For holographic CFTs in an appropriate semiclassical regime, the same kinematic geometry
acquires the usual dynamical interpretation as a bulk spacetime.

In this sense, the phrase ``antipodal geodesic on the EWCS'' refers first to the geometric
object determined by the solid torus conformal class. The terminology agrees with the
standard entanglement wedge picture when the CFT admits a semiclassical holographic dual,
but the correlator/geodesic pair itself only uses conformal kinematics. 
Similarly,  the bulk-bulk propagator should be understood in the fixed AdS kinematic geometry.
%the statement
%about the bulk-bulk propagator should be understood in the fixed AdS kinematic geometry:
%once the AdS dimension, the scalar dimension $\Delta$, and the boundary condition are fixed,
The scalar Green function is a function of the geodesic invariant by solving
\[
\left(-\nabla^2_{\rm AdS}+m^2\right)G_{\rm bb}^{(\Delta)}(X,X')
=
\frac{\delta(X,X')}{\sqrt{g}},
\quad
m^2=\Delta(\Delta-D).
\]
Our construction reconstructs precisely this invariant from the Weyl frame CFT
correlator.

%The geometric origin of Eq.~\eqref{eq:main-pairing-general} can be seen by writing the bulk metric in coordinates adapted to the Weyl frame,
%\begin{equation}
%ds^2_{\rm bulk}
%=
%[\,\text{insert Euclidean bulk metric adapted to the solid-torus/Weyl description}\,],
%\end{equation}
%for which the relevant geodesic lies at
%\begin{equation}
%[\,\text{insert location / turning point / symmetry condition}\,].
%\end{equation}
%Its proper length is then
%\begin{equation}
%\ell_{\rm bulk}(x_1,x_2)
%=
%\int_{\gamma(x_1,x_2)} ds
%=
%[\,\text{insert explicit result}\,].
%\label{eq:geodesic-length}
%\end{equation}
%Combining Eqs.~\eqref{eq:Weyl-transform} and \eqref{eq:geodesic-length} yields the exact pairing.

\paragraph*{Boundary-anchored limit.}
The familiar boundary-anchored geodesic relation is recovered as a singular boundary 
limit of the exact  pair.
This is most transparent in the 
cavity configuration as shown in Fig. \ref{fig:limit}.

In this cavity configuration, $|x_P - x_Q|=2R+2\epsilon$.
The dual bulk geodesic is the purple curve $\ell(p,q)$.
As $\epsilon\to 0$, the  RT surfaces (blue curves) collapse
and then $\ell(p,q)\to \infty$.
To calculate the distance specifically, 
we should use the conformal 
invariant expression (\ref{eq:GL-rho}) since eq. (\ref{eq:vGL}) only applies
to the symmetric solid torus in Fig. \ref{fig:solid-torus}. Using eq.  (\ref{eq:inversive product}),
we have $\varrho = \frac{R^2+\epsilon^2}{R^2-\epsilon^2}$ and then 
$\ell(p,q)= \log\frac{4R^2}{\epsilon^2} + {\mathcal O}(\epsilon^2)$. 
It is straightforward to verify
\begin{eqnarray}
\lim_{\epsilon\to 0} \epsilon^{-2\Delta} G_{\mathcal W}(P,Q) 
&=& \lim_{\epsilon\to 0} C_\Delta e^{-\Delta (\ell(p,q) + \log \epsilon^2)}\nonumber\\
&=& \frac{C_\Delta}{|x_P - x_Q|^{2\Delta}},
\end{eqnarray}
which is the familiar 
traditional boundary-anchored geodesic relation in the asymptotic limit,
where usually the regularized 
geodesic $\ell_\mathrm{reg}= \ell + \log\epsilon^2$ is adopted.
Thus the standard asymptotic dictionary arises as the boundary 
singular limit of the present exact
deep-bulk relation.

\paragraph*{Free scalar and mode expansion.}
The exact pairing admits an illuminating derivation for the free scalar. 
In the Weyl frame $S^1\times H^{D-1}$,
consider a conformally coupled free  scalar CFT$_D$ living in the 
symmetric open solid torus Fig. \ref{fig:solid-torus}, 
\begin{equation}
S= \int_0^{2\pi} d\theta \int_{H^{D-1}} d^{D-1}V_H \Big( |\partial_\theta \Phi|^2 
+|\nabla_H \Phi|^2 -\Delta^2 |\Phi|^2\Big),
\end{equation}
where $\Delta^2 = \xi R_H = \frac{(D-2)^2}{4}$ is the conformal coupling. 
$\nabla_H$  is the hyperbolic Laplacian.
Expanding Fourier modes along $\theta$, 
$\Phi(\theta,\mathbf{x}) = \frac{1}{\sqrt{2\pi}}\sum_{n=-\infty}^\infty e^{in\theta}\phi_n(\mathbf{x}_H)$,
we get
\begin{equation}
S= \sum_{n=-\infty}^\infty
\int_{H^{D-1}} d^{D-1}V_H \Big( 
|\nabla_H \phi_n|^2 +(n^2-\Delta^2) |\phi_n|^2\Big).
\end{equation}
Every mode $\phi_n$ is a massive scalar with mass $m^2=n^2-\Delta^2$.
The correlator is 
\begin{eqnarray}
G^{(D)}_{\mathcal W}(P,Q)&=& 
\langle \Phi(P)\Phi(Q)\rangle_{\mathcal W}^{(D)}\nonumber\\
&=&\frac{1}{2\pi} \sum_{n=-\infty}^\infty {e^{in\delta \theta}} G_{H^{(D-1)}}(\sigma;n^2),
\label{eq:Green}
\end{eqnarray}
where $\sigma$ is the hyperbolic distance between $P$ and $Q$ in $H^{(D-1)}$.
$G_{H^{(D-1)}}(\sigma;n^2)$ is the Green function of a massive scalar in $H^{D-1}$, 
\begin{eqnarray}
&&G_{H^{(D-1)}}(\sigma;n^2) = \frac{\Gamma(\Delta+|n|)}{2 \pi^\Delta \Gamma(|n|+1)}
\left(2\sinh\frac{\sigma}{2}\right)^{-2(\Delta+|n|)}\nonumber\\ 
&&\quad\quad\times\,_2F_1 \Big(\Delta +|n|, |n|+1;2|n|+1; -\sinh^{-2}\frac{\sigma}{2} \Big),
\label{eq:CompG}
\end{eqnarray}
where $_2F_1$ is the standard hypergeometric function. To compute the summation,
we use  the heat kernel
$G_{H^{(D-1)}}(\sigma;n^2) = \int^\infty_0 dt e^{-n^2 t} K_{H^{D-1}} (t;\sigma)$, with the  recursion
$ (2\pi \sinh\sigma) K_{H^{D+2}} (t;\sigma) = -\partial_\sigma K_{H^{D}} (t;\sigma)$. 
%\begin{equation}
%K_{H^{D+2}} (t;\sigma) = -\frac{1}{2\pi \sinh\sigma}\partial_\sigma K_{H^{D}} (t;\sigma),
%\end{equation}
This leads to a recursive relation for the correlator,
\begin{equation}
G_{\mathcal W}^{(D+2)} (\sigma,\delta\theta) = -\frac{1}{2\pi \sinh\sigma}\partial_\sigma 
G_{\mathcal W}^{(D)} (\sigma,\delta\theta).
\label{eq:recursion}
\end{equation}
For $D=3,4$, the sum can be evaluated explicitly,
\begin{eqnarray}
G_{\mathcal W}^{(3)} (\sigma,\delta\theta) &=& \frac{1}{4\pi}\frac{1}{\sqrt{2(\cosh\sigma-\cos\delta\theta)}},\\
G_{\mathcal W}^{(4)} (\sigma,\delta\theta) &=& \frac{1}{8\pi^2} \frac{1}{\cosh\sigma -\cos\delta\theta}.
\end{eqnarray}
Therefore,  using the recursion (\ref{eq:recursion}),
the intricate expression in  (\ref{eq:Green}) finally collapses to
\begin{equation}
G_{\mathcal W}^{(D)}(\sigma,\delta\theta) = \frac{\Gamma(\Delta)}{4\pi^{\Delta+1}} 
\frac{1}{[2(\cosh\sigma - \cos\delta\theta)]^\Delta},\quad \Delta = \frac{D-2}{2},
\end{equation}
which up to a normalization number, is precisely the correlator eq. (\ref{eq:WithTheta}) we obtained by symmetry approach.

%
%
%\begin{widetext}
%\[\bigoplus_{n\in \mathbb Z}\hbox{massive modes on }H^{(D-1)}\longrightarrow \hbox{CFT correlator on }
%S^1\times H^{(D-1)} \longleftrightarrow \hbox{finite geometric object in }AdS_{D+1}
%\]
%\end{widetext}

A central and genuinely nontrivial consequence of our construction is that the KK tower does not merely reconstruct the boundary two point function on \(S^1 \times H_{D-1}\). Rather, the full tower of $(D-1)$-dim (non-conformal) massive Euclidean resolvents resums exactly into a $D-$dim {\it conformal}  correlator, which in turn exactly matches 
a finite geodesic quantity in  AdS\(_{D+1}\) and then  reconstructs the exact bulk-bulk propagator  $G^{(\Delta)}_{\rm bb}(X,X')$ on AdS$_{D+1}$:
\begin{widetext}
\[
\hbox{massive resolvents on }H_{D-1}\to G^{(D)}_{\rm CFT}\, {\rm on}\, (S^1\times H_{D-1} )
\leftrightarrow \hbox{geodesics in AdS}_{D+1} \to G^{(\Delta)}_{\rm bb}(X,X').
\]
\end{widetext}
In this sense, our result goes well beyond an ordinary mode expansion or a formal resummation identity: an infinite collection of lower-dimensional non-conformal massive correlators reorganizes exactly into a single higher-dimensional geodesic and propagator structure. The nontrivial content is therefore not simply the existence of a closed-form sum, but the exact uplift from spectral data on \(H_{D-1}\) to the bulk geometric and propagator data on AdS\(_{D+1}\).

Conceptually, this also provides an explicit mechanism by which higher-dimensional conformal structure emerges from a highly organized tower of lower-dimensional massive QFT correlators, and vice versa.

\paragraph*{Conclusion and discussion.}
We have shown that for a CFT on a flat open solid torus, the Weyl frame correlator
is exactly paired with a finite geodesic in the  bulk interior.
In the free scalar case, the same result follows from a mode expansion 
in which the distinguished $S^1$ generates effective masses whose complicated propagator 
sum collapses to the same simple geodesic expression.

The exact pair identified here expresses the boundary quantity exactly 
in terms of a bulk geometric invariant, without invoking large \(N\), large \(\Delta\), or any saddle-point approximation. 
The standard boundary-anchored relation is a singular limit of this exact pair.

\paragraph*{Toward an exact-pair program.}
Taken together the exact pair ``disjoint entanglement entropy $\equiv$ EWCS'' built in
\cite{Jiang:2025jnk}, our construction suggests a broader exact-pair program in AdS/CFT:
rather than studying the duality only through asymptotic, semiclassical, or heavy operator correspondences, 
one may search systematically the exact pairs and treat them as fundamental objects.
%The present example indicates that ordinary two point functions on suitably chosen geometries can already realize this idea in a sharp and computable form.

As we can see, {\it the open solid torus provides a scale $r=R_1, R_2$ to 
supplement the planar coordinates; the 
Weyl frame manifests this scale which eventually represents the bulk 
direction. This is why (solid torus $+$ Weyl) leads to exact pairs.} 
%Several questions naturally follow. 
%What general conditions on the boundary geometry permit exact pairing?
%Can analogous relations be found for higher point functions, spinning operators, or nonconformal theories?
%And to what extent do these Euclidean exact pairs admit a controlled analytic continuation to Minkowsi?
%We hope the present example provides a useful starting point for addressing these questions.

%More broadly, 
%since the geodesic distance defines Synge's world function \cite{Synge:1960ueh}, 
%whose coincidence limit determines the local bulk metric as we did in 
%\cite{Jiang:2024hjz, Jiang:2024xqz, Jiang:2025lkx},
%this raises the possibility that boundary correlators directly encode 
%the background geometry itself. 
%From this perspective, the exact pair is best viewed as a kinematical 
%reorganization of CFT data into bulk geometry.

\vspace*{3.0ex}
\begin{acknowledgments}
\paragraph*{Acknowledgments.} 
We are indebted to Bo Feng and  Xin Gao for very useful comments which helped us to improve the manuscript substantially. 
This work is supported in part by NSFC (Grant No. 12275183, 12275184).
\end{acknowledgments}

\bibliographystyle{unsrturl}
\bibliography{ref202605}
%\onecolumngrid

\end{document}